\documentclass{article}

\usepackage[english]{babel}

\usepackage[letterpaper,top=2cm,bottom=2cm,left=3cm,right=3cm,marginparwidth=1.75cm]{geometry}

\usepackage{amsmath}
\usepackage{graphicx}
\usepackage[colorlinks=true, allcolors=blue]{hyperref}

\title{Virial clouds explaining the observed rotational asymmetry in the galactic halos}

\author{Asghar Qadir$^{1,*}$, Noraiz Tahir$^{1,\dagger}$, and Muhammad Sakhi$^{1,\ddagger}$ \\
$^1$Department of Physics, School of Natural Sciences\\
National University of Sciences and Technology, H-12, Islamabad, Pakistan\\
$^*$asgharqadir46@gmail.com \\ $^\dagger$ noraiztahir78637@gmail.com \\ $^\ddagger$ muhammad.saki@yahoo.com}

\begin{document}
\maketitle

\begin{abstract}
Rotation of galactic objects has been seen in the CMB that could be ascribed to molecular hydrogen clouds with, or without, dust contamination and contamination from other sources. We model the clouds using the canonical ensemble for pure molecular hydrogen, a mixture of hydrogen helium and or dust, in order to constrain the physical parameters of these clouds. Since, the clouds are cold, we justify the use of the canonical ensemble by explicitly calculating the interaction between the hydrogen molecules and the CMB photons and determining the time required for thermal equilibrium to be reached, and show that there is enough time for the equilibrium to be attained.
\end{abstract}

\section{\label{intro}Introduction}
The Universe contains 5\% baryons, only half of which are visible \cite{1,2}. The question of the missing baryons is still unanswered. Different scenarios have been proposed to answer this question \cite{3,4,5,6,7,8,9}. Nearly, 25 years ago, it was proposed that some small fraction of primordial hydrogen not swept away during star formation, would have started to collapse due to gravitational instability, and form clouds which contribute to the galactic halo dark matter \cite{10}. These clouds would then radiate, until they reach, the cosmic microwave background radiation (CMB) temperature, after which they would not be able to radiate further and become stable. As such we call them ``virial clouds''. The question is, how can we see such clouds that would exactly merge into the (cosmic) background like chameleons.

There were two proposals \cite{3}: (a) look for $\gamma$-ray scintillation due to cosmic rays striking these clouds; (b) look for Doppler shifts in the CMB as seen through M31. Though the $\gamma$-rays were seen, they had too many possible sources \cite{11,12,13}. The clouds in the halo of M31 rotating towards us would be blue-shifted and the part moving away would be red-shifted. WMAP data of 2011 showed the effect, at a low but significant level \cite{14}. Later, Planck data of 2014 confirmed the effect in the M31 halo at a sufficient accuracy, to provide convincing evidence which was further used to map the rotation of the M31 galaxy \cite{15}. The data was then used to map the rotational dynamics and the effect in various galaxies, including, NGC 5128, M33, M81, and M82 \cite{16,17,18,19}. 

The verification of the prediction based on the proposed hydrogen clouds does not prove that it is caused by those clouds. To be able to test the proposal we need to model hydrogen clouds, hydrogen-helium clouds, and compare the results with (say) dust clouds; a mixture of hydrogen and dust; and a mixture of hydrogen, helium and dust. Since the proposed clouds are ``held up'' by the CMB, they are in a heat bath and so should be modeled by an isothermal gas sphere. However, this would require that these cold clouds can interact with the low temperature CMB. To check this we calculate the interaction between the CMB and the hydrogen molecules to determine the time it would take to reach thermal equilibrium.

The plan of the paper is as follows: In Section \ref{virialmodel} we will assume that these clouds contain only molecular hydrogen or only interstellar dust or only helium. Then we try to estimate the physical parameters of these considered clouds. In Section \ref{twofluid} we consider the two-fluid model, and use the above procedure to try to estimate how the cloud's physical parameters vary as a function of the ratio between molecular hydrogen and dust, and go on to model the cloud's parameters with a fixed fraction of hydrogen (75\%) and helium (25\%). In Section \ref{threefluid} we will extend our analysis to a three-fluid model, and model the possibility of clouds containing hydrogen, helium and dust with varying fractions. In Section \ref{probability} we estimate the stability time for the clouds to reach equilibrium and merge with the CMB. It is shown that there is adequate time before collapse for stability. At the end in Section \ref{conclusions} the results and conclusions will be discussed.
\section{\label{virialmodel} The Virial Model}
As the virial clouds are formed due to the Jeans instability, we need to replicate the Jeans analysis \cite{20} for our purpose. He used the Lane-Emden equation for an isothermal gas sphere and required that at the boundary the density merges with the ISM. One needs to then use some density distribution on an ad-hoc basis. We had used it to model the distribution of these clouds in the halo of M31 \cite{21} and several other nearby spiral galaxies \cite{ajom}. Here we will start from scratch, assuming that the clouds are exactly at the CMB temperature, so they are immersed in a heat bath which is the CMB and are completely merged with it. Hence we need to use the canonical distribution, and not the microcanonical \cite{padmanaman}. We require that the density profile be flat at the center and zero at the boundary. In this case the equation remains the same but we get different boundary conditions and avoid the ``Jeans fiddle''. The advantage of starting from scratch is that we can incorporate contamination of the hydrogen cloud by dust; hydrogen cloud by helium; and then hydrogen-helium cloud by dust.
\subsection{Pure Molecular Hydrogen Clouds}
\begin{figure}[htbp]
  \centering
  \includegraphics[width=9.3cm]{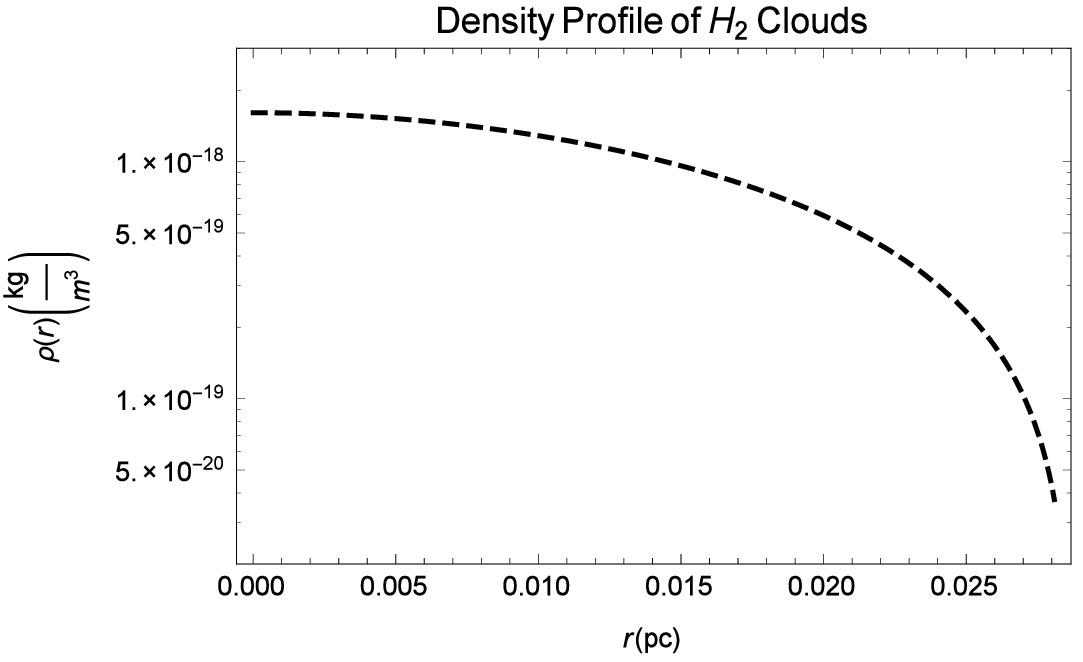}
\caption{The curve represents the density of the virial clouds, assuming that they are only composed of molecular hydrogen. It is clear from the figure that the central density of the cloud is ${\displaystyle \rho_{c} \simeq 1.60 \times 10^{-18}~{\rm kgm^{-3}}}$. The density goes on decreasing and is exactly zero at $R_J \simeq 0.032~{\rm pc}$.}
\label{f1}
\end{figure}

Let us consider a spherical cloud of pure molecular hydrogen, each molecule of hydrogen has a mass $m_H\approx 3.35 \times 10^{-27}~{\rm kg}$, and the total mass of that cloud is $M_H$. The cloud is allowed to collapse due to the gravitational instability. From the virial theorem, $2K+\Phi=0$, the Jeans mass squared is \cite{22}
\begin{equation}
M_{J}^2 \simeq \left(\frac{81}{32\pi\rho_{c}}\right)\left(\frac{3c_{s}^2}{5G}\right)^3,
\label{1}
\end{equation}
and the corresponding Jeans length (radius) squared is
\begin{equation}
R_{J}^{2}=\frac{27c_s^2}{20\pi\rho_{c}G},
\label{2}
\end{equation} 
where, $\rho_{c}$ is the central density, $G$ is Newton's gravitational constant and $c_{s}$ is the isothermal speed of sound. This speed is given in terms of the mass of the molecules, $m$, and the temperature of the cloud, $T$. For our purposes $T=T_{CMB}\approx2.7254 {\rm ^{0} K}$, is directly related to the mass of the molecules present inside the cloud, and the average speed. So,
\begin{equation}
c_{s}=\sqrt{\frac{\gamma kT_{CMB}}{m}},	
\label{3}
\end{equation}
where $k$ is the Boltzmann constant, and $m=m_H$. Since even the vibrational and the rotational modes are not excited at the CMB temperature, so they don't vibrate nor rotate at it. Hence, $\displaystyle{\gamma=5/3}$, which is the ideal gas approximation. So, $c_{s}\simeq 1.36 \times 10^{2}~{\rm m~s^{-1}}$. The canonical ensemble distribution is given by the relation (see \cite{23})
\begin{equation}
f(r,p)=\frac{1}{h^{3N}N!}\frac{1}{Z}exp\left(-\frac{H(r,p)}{k~T_{CMB}}\right),
\label{e4}
\end{equation}
where, $H(r,p)$ is the Hamiltonian of the system, and $Z$ is the partition function which is defined as
\begin{equation}
Z=\frac{1}{h^{3N}N!}\int_{V}\int_{V} d^{3}p~d^{3}r~exp\left(\frac{-H(r,p)}{k~T_{CMB}}\right).
\label{5}
\end{equation}
The Hamiltonian is given by
\begin{equation}
H(r,p)=p^2/2m+GM(r)m/r.
\label{6}
\end{equation}
where, $p=mc_{s}$ is the momentum of the molecules. From eqs. (\ref{5}) and (\ref{6}) we have the partition function given as
\begin{equation}
Z=\frac{1}{h^{3N}N!}\frac{3^{3/2}\pi^{3/2}}{2^{1/2}}\frac{(kT_{CMB})^3}{(G\rho_c)^{3/2}}.
\label{extra1}
\end{equation}
Eq. (\ref{e4}) will take the form 
\begin{align}
f(r,p)=&\frac{2^{1/2}}{3^{3/2}\pi^{3/2}}\frac{(G\rho_c)^{3/2}}{(kT_{CMB})^3}\nonumber \\ &exp\left(-\left[\frac{p^2}{2mkT_{CMB}}+\frac{3GM(r)}{5rc_{s}^2}\right]\right).
\label{extra2}
\end{align}
Now, 
\begin{equation}
\rho(r)=\int_{-\infty}^{\infty}4\pi mp^2 f(r,p) dp.
\label{extra3}
\end{equation}
So, we have the density distribution given by 
\begin{align}
\rho(r)=8m^{5/2}\left(\frac{G\rho_{c}}{3kT_{CMB}}\right)^{3/2}
exp\left(-\frac{3GM(r)}{5rc_{s}^2}\right),	
\label{7}
\end{align}
where, $M(r)$ is the total mass of the cloud interior to $r$ and is defined as
\begin{eqnarray}
M(r)=\int_{0}^{r}4\pi\rho(q)q^{2}~dq.
\label{8}
\end{eqnarray}
The boundary conditions are that the central density is $\rho_c$ and $(d\rho(r)/dr)|_{r\rightarrow0}=0$. Taking the natural logarithm of eq. (\ref{7}) and substituting eq. (\ref{8}) we get 
\begin{align}
\int_{0}^{r}4\pi q^{2}\rho(q)~dr=-\left(\frac{rk~T_{CMB}}{4\pi mG}\right)\ln\left(\frac{\rho(r)}{\zeta}\right),
\label{9}
\end{align}
where, $\zeta=(8m^{5/2}/3^{3/2})(G\rho_c/k~T_{CMB})^{3/2}$. Taking the derivative of eq. (\ref{7}) with respect to $r$, and substituting eq. (\ref{9}). The differential equation is given as
\begin{align}
r\frac{d\rho(r)}{dr}-r^{2}\left(\frac{4\pi Gm}{k~T_{CMB}}\right)\rho^{2}(r)-\rho(r)\ln\left(\frac{\rho(r)}{\zeta}\right)=0.
\label{eq17}
\end{align}
This is essentially the Lane-Emden equation. We now solve it numerically with a guess value for the central density and see where the density becomes zero. We then check the value of the Jeans radius with that central density. We then adjust the value of the central density so that the density becomes zero exactly at the Jeans radius. In this way we get a self-consistent solution of the differential equation subject to the given boundary conditions. The result of the calculation, depicted in Fig.\ref{f1}, yield $\displaystyle{\rho_{c}\simeq 1.60 \times 10^{-18}~{\rm kgm^{-3}}}$, the Jeans Radius, $R_{J}\simeq 0.032~{\rm pc}$, and the Jeans mass is $M_{J}\simeq 0.78~M_{\odot}$.
\subsection{Pure Interstellar Dust Clouds}
We have estimated physical parameters of the virial clouds, with the assumption that they are pure molecular hydrogen clouds. There is also a possibility that the left over interstellar dust will form another cloud without the contamination of molecular hydrogen. 

Assuming that a single dust grain is ${\rm CN_2O_3}$, then the mass of a single grain will then be $\displaystyle{m_d\approx1.46 \times 10^{-25}{\rm kg}}$, the total mass of this cloud interior to $r$. The mass density profile, of these clouds, is shown Fig.\ref{f2}. The central density for the dust cloud is ${\displaystyle \rho_c \simeq 1.46 \times 10^{-17}~{\rm kgm^{-3}}}$, the Jeans mass is $\displaystyle{M_J=8.98 \times 10^{-4}~M_\odot}$, and the corresponding Jeans radius, $R_J \simeq 1.40 \times 10^{-3}~{\rm pc}$. It is clear from Fig.\ref{f1} and Fig.\ref{f2} that the density of the dust cloud is greater than that of pure molecular hydrogen cloud but the Jeans radius is smaller.
\begin{figure}[htbp]
  \centering
  \includegraphics[width=9.3cm]{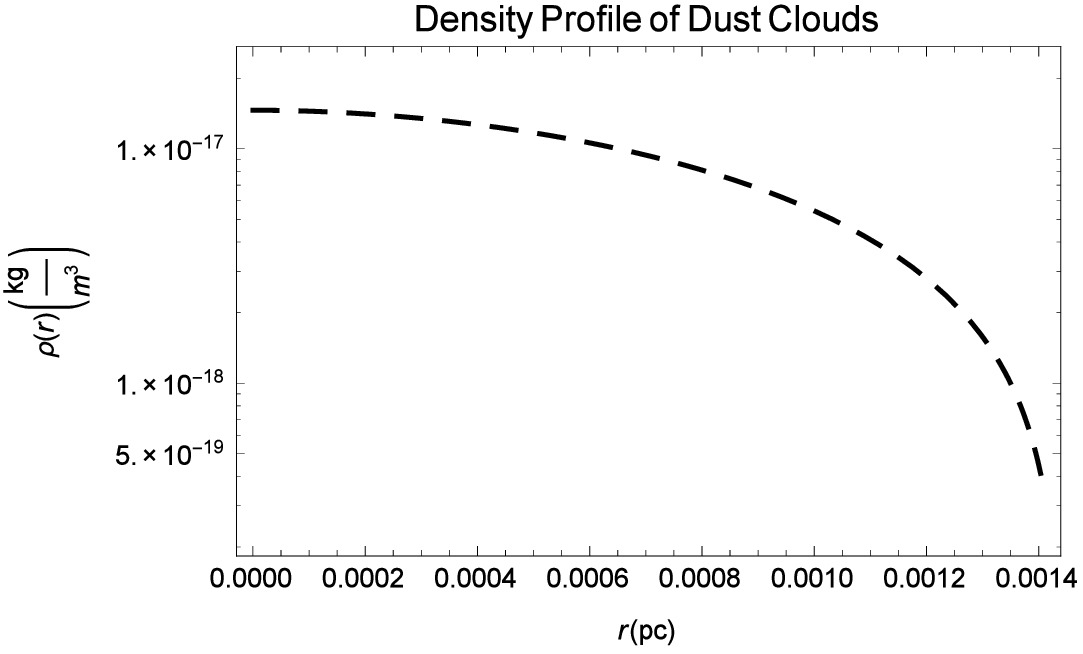}
\caption{The curve represents the density of the dust clouds. It is clear from the figure that the central density of the cloud is $\displaystyle{\rho_{c}\simeq 1.46 \times 10^{-17}~{\rm kgm^{-3}}}$. The density goes on decreasing and is zero at $R_J \simeq 1.40 \times 10^{-3}~{\rm pc}$.}
\label{f2}
\end{figure}
\subsection{Pure Helium Clouds}
We have now estimated the physical parameters of pure molecular hydrogen cloud and pure dust cloud. It is also possible that the left-over helium ($He$) present in ISM form clouds that are at the CMB temperature, without the contamination of dust or molecular hydrogen, so, we need to model this possibility also. The mass of single helium molecule is $m_{He}= 6.64 \times 10^{-27}~{\rm kg}$. Using the same procedure as we did for molecular hydrogen cloud and dust cloud, we have the density profile for a pure helium cloud is shown in Fig.\ref{helium}.  The central density for the helium cloud is estimated to be ${\displaystyle \rho_c \simeq 6.98 \times 10^{-18}~{\rm kgm^{-3}}}$, the Jeans mass is $\displaystyle{M_J=0.13~M_\odot}$, and the corresponding Jeans radius, $R_J \simeq 0.013~{\rm pc}$. It is clear from Fig.\ref{f1}, Fig.\ref{f2} and Fig.\ref{helium} that the density of the dust cloud is greater than that of pure helium cloud, and pure molecular hydrogen cloud but the Jeans radius is smaller.
\begin{figure}[htbp]
  \centering
  \includegraphics[width=9.3cm]{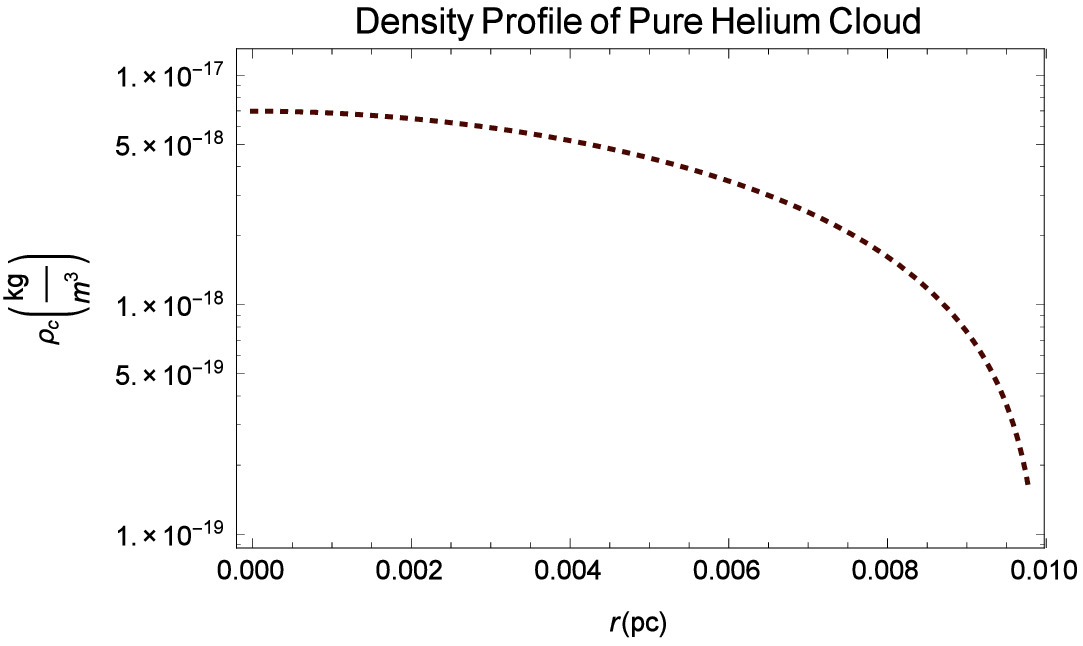}
\caption{The curve represents the density of helium clouds. It is clear from the figure that the central density of the cloud is $\displaystyle{\rho_{c}\simeq 6.98 \times 10^{-18}~{\rm kgm^{-3}}}$. The density goes on decreasing and is zero at $R_J \simeq 0.013~{\rm pc}$.}
\label{f3}
\end{figure}
\section{\label{twofluid}The Two Fluid Model}
In this section we model the virial clouds as a mixture of two components: hydrogen and interstellar dust; and hydrogen and helium. We will try to estimate the change in the physical parameters of virial clouds for different fractions of molecular hydrogen and dust (see Section \ref{hdc}), and then  for the contamination of helium, we will fix the fraction to 75\% hydrogen and 25\% of helium (see Section \ref{hhec}) and check what changes might occur in the mass and radius of the modeled clouds.       
\subsection{Hydrogen-Dust Clouds \label{hdc}}
As explained earlier,we need to extend our analysis to a mixture of two fluids, hydrogen and dust. The total mass of the cloud will be $M_{cl}(r)=\alpha~M_{H}(r)+\beta~M_{d}(r)$, where, $M_H(r)$ is the total mass of the molecules of hydrogen, and $M_d(r)$ is the total mass of the dust grains interior to $r$, where, $\alpha$ and $\beta$ are the fractions of hydrogen and dust, so that $\alpha+\beta=1$
. Let the mass density of the whole cloud is $\rho_{cl}$. We have two fluids whose molecules are distinguishable and non reactive. The partition function of the system will be $Z=Z_H.Z_d$, where, $Z_H$ is for hydrogen molecules and $Z_d$ for dust molecules. Similarly, the canonical ensemble distribution will become
\begin{align}
f(r,p)=\frac{1}{h^{(3N_H+3N_d)}N_H!N_d!}\frac{1}{Z}exp\left(-\left[\frac{H_H(r,p_H)}{kT_{CMB}}+\frac{H_d(r,p_d)}{kT_{CMB}}\right]\right),
\end{align}
where, $N_H$ and $N_d$ are total number of molecules of hydrogen and dust, $H_H(r,p_H)$ and $H_d(r,p_d)$ are the Hamiltonian for molecular hydrogen and dust grain clouds. The mass density distribution of the two fluid model is given by
\begin{align}
\rho_{cl}(r)=\sqrt{\frac{64}{27}}\frac{(G\rho_{c_H}\rho_{c_d})^{3/2}}{(kT_{CMB})^{9/2}}(m_Hm_d)^{5/2}exp\left[-\frac{1}{2}\left(\frac{\alpha G M_H(r)m_H}{rkT_{CMB}}+\frac{\beta G M_d(r)m_d}{rkT_{CMB}}\right)\right],
\end{align}
where, $\rho_{c_H}$ and $\rho_{c_d}$ are the central density of the molecular hydrogen and dust cloud. Also
\begin{align}
\int_{0}^{r}(\alpha m_H\rho_H(q)+\beta m_d\rho_d(q))q^2~dq=-\left(\frac{2rkT_{CMB}}{4\pi G}\right)\ln\left(\frac{\rho_{cl}(r)}{\eta}\right),
\end{align}
and $\eta=(64/27)^{1/2}[(G\rho_{c_H}\rho_{c_d})^{3/2}/(kT_{CMB})^{9/2}][m_Hm_d]^{5/2}$. The differential equation for the two fluid model is
\begin{align}
r\frac{d\rho_{cl}(r)}{dr}-r^2\left(\frac{2\pi G}{kT_{CMB}}\right)[\rho_{cl}(r)(\alpha\rho_{c_H}m_H+\beta\rho_{c_d}m_d)]-\rho_{cl}(r)\ln\left(\frac{\rho_{cl}(r)}{\eta}\right)=0.
\end{align}
The density distribution, with different fractions of hydrogen and dust, is shown in Fig.\ref{f3}. It is seen that with the increase in contamination of dust, the density of virial cloud increases, but the Jeans radius is decreased and so is the corresponding Jeans mass.
\begin{figure}[htbp]
  \centering
  \includegraphics[width=9.3cm]{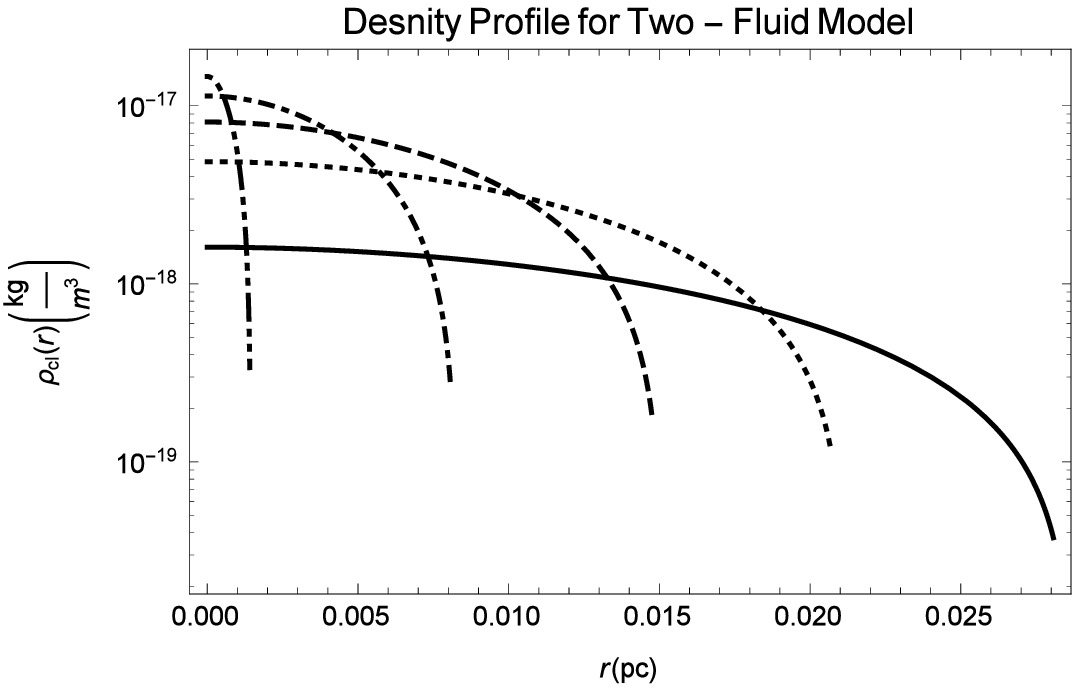}
\caption{The curves in this figure represents the density profile for different values of $\alpha$ and $\beta$. The bold black curve represents the density profile when $\alpha=1$ and $\beta=0$, dotted represents the density profile when $\alpha=0.75$ and $\beta=0.25$, dashed represents the density profile when $\alpha=0.5$ and $\beta=0.5$, dotted-dashed represents the density profile when $\alpha=0.25$ and $\beta=0.75$, and dot-dot-dashed represents the density profile when $\alpha=0$ and $\beta=1$. It is clearly seen that the central density $\rho_{c}$ is different for different concentrations of molecular hydrogen and interstellar dust. The density of the virial cloud increases with the increase in the concentration of interstellar dust in the cloud and the size of the cloud is decreased.}
\label{f2}
\end{figure}
\begin{table}[htb]
 \centering \caption{Physical parameters for the considered two-fluid hydrogen and dust model}
 \label{t1}
\begin{tabular}{c c c c c}
    \hline
    	$\alpha$ & $\beta$ & Central Density  & Jeans mass & Jeans Radius\\
		&     & $\rho_{c} ({\rm kg~m^{-3}}) $& $M_{\odot}$ & pc\\    
		\hline  
		1 & 0 & $1.60 \times 10^{-18}$   &0.78 & 0.032\\ 
		0.75&0.25&  $5.00 \times 10^{-18}$ & 0.58& 0.021\\
		0.5& 0.5 & $6.58 \times 10^{-18}$  & 0.39& 0.015\\
		0.25& 0.75 & $1.00 \times 10^{-17}$ &0.19& 0.008\\
		0& 1 & $1.46 \times 10^{-17}$   & 0.00089& 0.0014\\
    \hline
   \end{tabular}
   
   {\bf Note:} We give the central density $\rho_c$ (column 3), for different fractions of molecular hydrogen and interstellar dust and the corresponding Jeans mass (column 4), and Jeans radius (column 5) of the virial clouds.
    \end{table}
\subsection{Hydrogen-Helium Clouds \label{hhec}}
\begin{figure}[htbp]
  \centering
  \includegraphics[width=9.3cm]{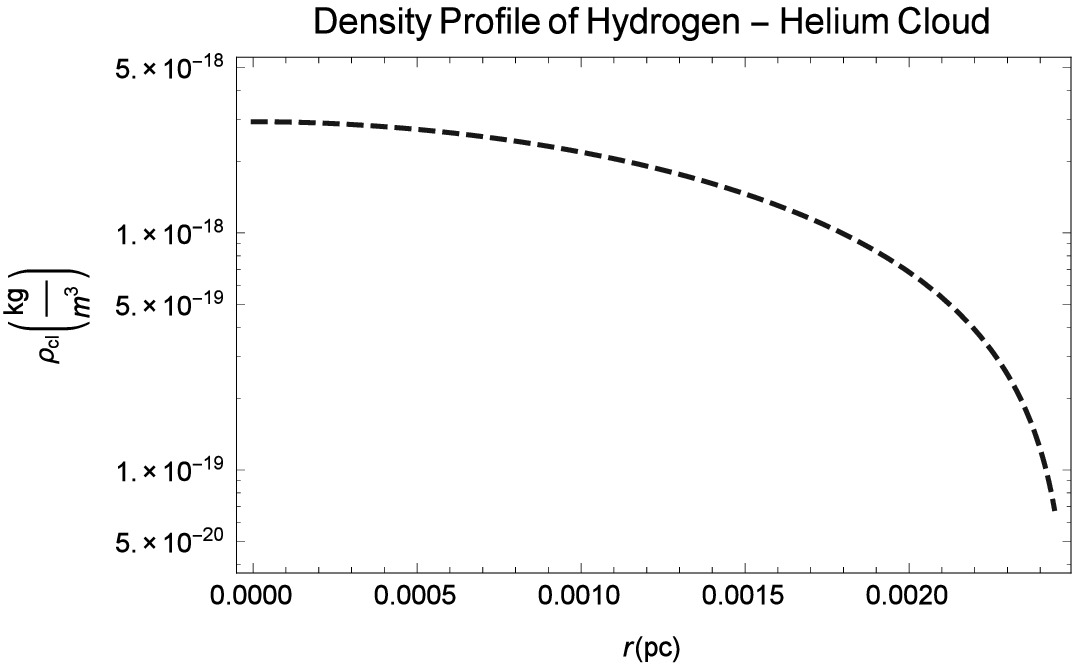}
\caption{The curve in this figure represents the density profile for hydrogen-helium two fluid model. It is clearly seen that the central density is $\rho_{c}=4.87 \times 10^{-18}~{\rm kg~m^{-3}}$, and it goes exactly to zero at $r=R_J=0.025$ pc.}
\label{fhhe}
\end{figure}

Since, primordially ISM not only contain hydrogen but there was approximately 75\% of hydrogen and 25\% of helium. So, it is possible that the clouds are not pure molecular hydrogen clouds there is 25\% contamination of helium in them. Now the total mass of the cloud for this case will be $M_{cl}(r)=0.75~M_{H}(r)+0.25~M_{He}(r)$, where, $M_{He}(r)$ is the total mass of the helium interior to $r$. Eq.(15), (16) and (17) can be re-written as
\begin{align}
	\rho_{cl}(r)=\sqrt{\frac{64}{27}}\frac{(G\rho_{c_H}\rho_{c_{He}})^{3/2}}{(kT_{CMB})^{9/2}}(m_Hm_{He})^{5/2}exp\left[-\frac{1}{2}\left(\frac{0.75 G M_H(r)m_H}{rkT_{CMB}}+\frac{0.25 G M_d(r)m_{He}}{rkT_{CMB}}\right)\right],
\end{align}
where, $\rho_{c_H}$ and $\rho_{c_{He}}$ are the central density of the molecular hydrogen and helium cloud. Also
\begin{align}
	\int_{0}^{r}(0.75 m_H\rho_H(q)+0.25 m_{He}\rho_{He}(q))q^2~dq=-\left(\frac{2rkT_{CMB}}{4\pi G}\right)\ln\left(\frac{\rho_{cl}(r)}{\tau}\right),
\end{align}
and $\tau=(64/27)^{1/2}[(G\rho_{c_H}\rho_{c_{He}})^{3/2}/(kT_{CMB})^{9/2}][m_Hm_{He}]^{5/2}$. The differential equation for the two fluid hydrogen-helium model will be
\begin{align}
	r\frac{d\rho_{cl}(r)}{dr}-r^2\left(\frac{2\pi G}{kT_{CMB}}\right)[\rho_{cl}(r)(0.75\rho_{c_H}m_H+0.25\rho_{c_{He}}m_{He})]-\rho_{cl}(r)\ln\left(\frac{\rho_{cl}(r)}{\tau}\right)=0.
\end{align}
The density distribution of hydrogen-helium cloud, is shown in Fig.\ref{fhhe}. The obtained central density for this case is $\rho_{c}=4.87 \times 10^{-18}~{\rm kg~m^{-3}}$ and the corresponding Jeans mass $M_J=0.32 M_{\odot}$ and Jeans radius $R_J=0.025$ pc.
\section{Three Fluid Model\label{threefluid}}
\begin{figure}[htbp]
  \centering
  \includegraphics[width=9.3cm]{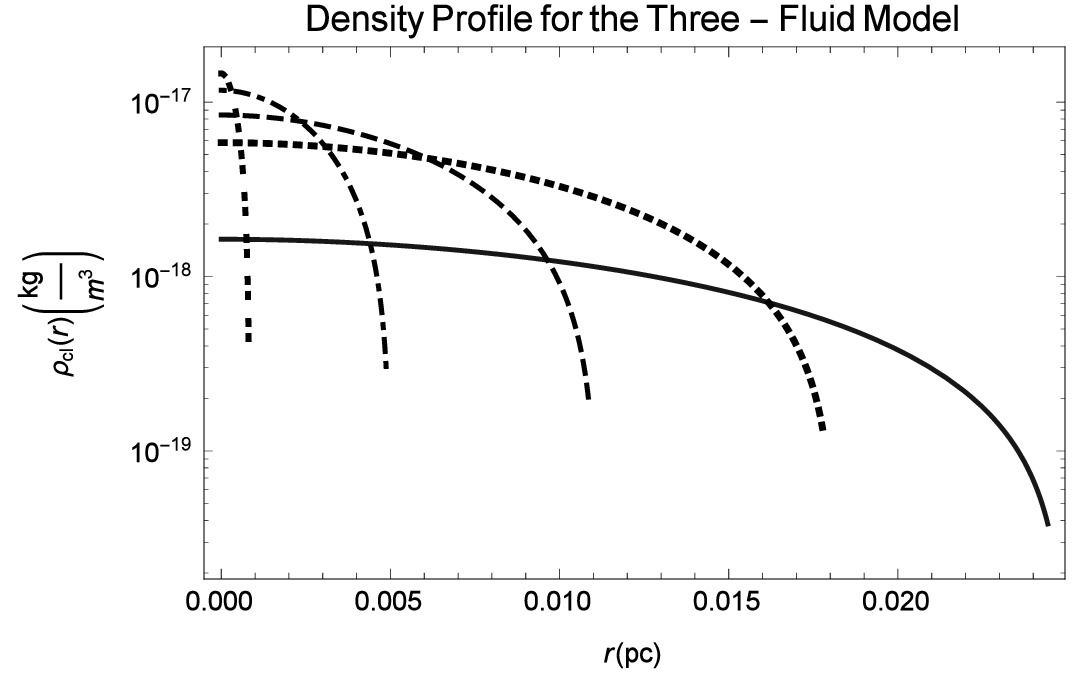}
\caption{The curves in this figure represents the density profile for different values of $\alpha$, $\beta$, and $\gamma$. The bold black curve represents the density profile when $\alpha=0.75$, $\beta=0$ and $\gamma=0.25$, dotted represents the density profile when $\alpha=0.56$, $\beta=0.25$, and $\gamma= 0.18$, dashed represents the density profile when $\alpha=0.37$, $\beta=0.50$, and $\gamma= 0.12$, dot-dashed represents the density profile when $\alpha=0.18$, $\beta=0.75$, and $\gamma= 0.06$, and dot-dot-dashed represents the density profile when $\alpha=0$, $\beta=1$, and $\gamma=0$. It is clearly seen that the central density $\rho_{c}$ is different for different concentrations of molecular hydrogen, interstellar dust, and helium. The density of the virial cloud increases with increase in the concentration of interstellar dust, and helium in the cloud and the size of the cloud is decreased. The results are not very different from those obtained for the two-fluid hydrogen and dust model.}
\label{f4}
\end{figure}

In this section we will model pure molecular hydrogen clouds with the contamination of helium, and interstellar dust in them, with the aim to estimate the variation in the physical parameters of the clouds. The total mass of the cloud will be $M_{cl}(r)=\alpha~M_{H}(r)+\beta~M_{d}(r)+\gamma M_{He}(r)$, where, $M_H(r)$ is the total mass of the molecules of hydrogen, $M_d(r)$ is the total mass of the dust grains, and $M_{He}(r)$ is the total mass of helium within $r$, where, $\alpha$, $\beta$ and $\gamma$ are the fractions of hydrogen, dust, and helium, so that $\alpha+\beta+\gamma=1$. Let the mass density of the whole cloud is $\rho_{cl}$. We have three fluids whose molecules are distinguishable and non reactive. The partition function of the system will be $Z=Z_H.Z_d.Z_{He}$, where, $Z_H$ is for hydrogen molecules, $Z_d$ for dust molecules, and $Z_{He}$ for helium. Similarly, the canonical ensemble distribution will become
\begin{align}
	f(r,p)=\frac{1}{h^{(3N_H+3N_d+3N_{He})}N_H!N_d!N_{He}!}\frac{1}{Z}exp\left(-\left[\frac{H_H(r,p_H)}{kT_{CMB}}+\frac{H_d(r,p_d)}{kT_{CMB}}+\frac{H_{He}(r,p_{He})}{kT_{CMB}}\right]\right),
	\label{threefluid1}
\end{align}
where, $N_H$, $N_d$, and $N_{He}$ are total number of molecules of hydrogen, dust, and helium, $H_H(r,p_H)$, $H_d(r,p_d)$, and $H_{He}(r,p_{He})$ are the Hamiltonian for molecular hydrogen, dust grain, and helium clouds. The partition function will become
\begin{align}
	Z=\frac{1}{h^{(3N_H+3N_d+3N_{He})}N_H!N_d!N_{He}!}\left(\frac{2^{1/2}3^{3/2}}{\pi^{1/2}}\right)\left(\frac{(kT_{CMB})^9}{G^{9/2}}\right)\left(\frac{1}{\rho_{c_H}\rho_{c_d}\rho_{c_{He}}}\right)^{3/2},
	\label{threefluid2}
\end{align}
where, $\rho_{c_H}$, $\rho_{c_d}$ and $\rho_{c_{He}}$ are the central densities of pure molecular hydrogen, dust and helium clouds. Substituting eq.(\ref{threefluid2}) in eq.(\ref{threefluid1}) we have the distribution as
\begin{align}
	f(r,p)=\left(\frac{\pi^{1/2}}{2^{1/2}3^{3/2}}\right)\left(\frac{G^{9/2}}{(kT_{CMB})^9}\right)\left(\rho_{c_H}\rho_{c_d}\rho_{c_{He}}\right)^{3/2}exp\left(-\left[\frac{H_H(r,p_H)}{kT_{CMB}}+\frac{H_d(r,p_d)}{kT_{CMB}}+\frac{H_{He}(r,p_{He})}{kT_{CMB}}\right]\right).
	\label{threefluid3}
\end{align}
Similarly, the density distribution can be found out by 
\begin{align}
(\rho_{cl}(r))^3=(16\pi^2)3\int_{0}^{\infty}p_H^2m_Hf(r,p_H)dp_H\int_{0}^{\infty}p_d^2m_df(r,p_d)dp_d~\int_{0}^{\infty}p_{He}^2m_{He}f(r,p_{He})dp_{He}.
\label{threefluid4}
\end{align}
Solving eq.(\ref{threefluid4}) we have the density distribution as
\begin{align}
	&\rho_{cl}(r)=\left(\frac{512\pi^{9/2}}{3^{9/2}}\right)\left(\frac{G}{kT_{CMB}}\right)^{9/2}(\rho_{c_H}\rho_{c_d}\rho_{c_{He}})^{3/2}(m_Hm_dm_{He})^{5/2}\nonumber \\ &exp\left[-\frac{1}{3}\left(\frac{\alpha GM_H(r)m_H}{rkT_{CMB}}+\frac{\beta GM_d(r)m_d}{rkT_{CMB}}+\frac{\gamma GM_{He}(r)m_{He}}{rkT_{CMB}}\right)\right].
	\label{threefluid5}
\end{align}
Similarly,
\begin{align}
\int_{0}^{r}(\alpha m_H\rho_H(q)+\beta m_d\rho_d(q)+\gamma m_{He}\rho_{He}(q))q^2~dq=-\left(\frac{3rkT_{CMB}}{4\pi G}\right)\ln\left(\frac{\rho_{cl}(r)}{\lambda}\right),
\label{threefluid6}
\end{align}
and $\lambda=(512 \pi^{9/2}/3^{9/2})[G/(kT_{CMB})]^{9/2}(\rho_{c_H}\rho_{c_d}\rho_{c_{He}})^{3/2} \newline[m_Hm_dm_{He}]^{5/2}$. The differential equation for the two fluid model is
\begin{align}
r\frac{d\rho_{cl}(r)}{dr}-r^2\left(\frac{2\pi G}{kT_{CMB}}\right)[\rho_{cl}(r)(\alpha\rho_{c_H}m_H+\beta\rho_{c_d}m_d+\gamma \rho_{c_{He}}m_{He})]-\rho_{cl}(r)\ln\left(\frac{\rho_{cl}(r)}{\lambda}\right)=0.
\label{threefluid7}
\end{align}
The density distribution, with different fractions of hydrogen, dust, and helium, by fixing 75\% concentration of molecular hydrogen, and 25\% of helium, is shown in Fig.\ref{f4}.
\begin{table}[htb]
 \centering \caption{Physical parameters for the considered three-fluid hydrogen, dust, and helium model}
 \label{t2}
\begin{tabular}{c c c c c c}
    \hline
    		$\alpha$ & $\beta$ & $\gamma$ & Central Density  & Jeans mass & Jeans Radius\\
		&     & & $\rho_{c} ( 10^{-18} {\rm kg~m^{-3}}) $& $M_{\odot}$ & pc\\    
		\hline  
		0.75 & 0 & 0.25 & 	$4.87$   &0.32 & 0.025\\ 
		0.5625 &0.25 & 0.1875 &  $5.60$ & 0.12& 0.018\\
		0.375 & 0.50 & 0.125 & $7.60$  & 0.089& 0.011\\
		0.1875 & 0.75 & 0.0625 & $14.4$ &0.056& 0.005\\
		0 & 1 & 0 & $14.6$   & 0.00089& 0.0014\\
    \hline
   \end{tabular}
   
   {\bf Note:} We give the central density $\rho_c$ (column 3), for different fractions of molecular hydrogen, interstellar dust, and helium, and the corresponding Jeans mass (column 4), and Jeans radius (column 5) of the virial clouds.
    \end{table}
    
\section{Stability of the Virial Cloud  \label{probability}}
One of problems associated with the existence of these clouds in the galactic halos is the that to explain the dynamical stability. The aim of this section is to estimate the scattering cross section and then using it to find the probability of interaction to estimate the stability time of the modeled virial clouds. We only estimated it for pure $H_2$ cloud, the other cases would not be substantially different. Though helium is monatomic, since there is no rotational or vibrational mode excited, the result will not be changed, except because of the size and density of the cloud. For dust it could be that some higher mode would be excited, but that would only make the probability of interaction with the CMB greater. We first treat the interaction as hard sphere collisions and will next give the quantum scattering. The CMB number density is $n_p=4.50~{\rm m^{-3}}$ and the estimated number of CMB photons that are interacting with the cloud are $N_p=1.79 \times 10^{57}$. The probability of interaction will be given by the relation
\begin{equation}
P=\frac{N\sigma_H}{dA},
\label{prob1}
\end{equation}
where, $N$ is the total number of $H_2$ molecules inside the cloud, and $\sigma_H$ is the cross-section area of each molecule. If $N_p$ is the total number of photons interacting with the cloud then eq.(\ref{prob1}) yields
\begin{equation}
P(r)=\int_{0}^{R_J}n_c\sigma_H exp\left(-n_c\sigma_H\right)~dr,
\label{prob}
\end{equation}
where, $n_c=4.78 \times 10^8~{\rm m^{-3}}$ is the number density of the pure $H_2$ cloud, $R_J$ is the Jeans radius for a pure $H_2$ cloud which is $0.032$ pc, and $\sigma_H$ is the cross section of a single molecule of hydrogen. Since, $\sigma_H=\pi r_H^2$, and $r_H=1.06 \times 10^{-9}$ m, the radius of single molecule of hydrogen, so, the cross section will be $\approx3.53 \times 10^{-20}~{\rm m^2}$. Solving eq.(\ref{prob}) we get the probability of interaction equal to $\approx 1~{\rm photon}/s$.

For the quantum calculation, since: (i) the energy of the CMB photon is $1.06 \times 10^{-22}$ J, and the estimated rest mass energy of $H_2$ molecule is $\approx 3.10 \times 10^{-10}$ J; (ii) the wavelength of the CMB photon is $\lambda_{CMB}=0.001$ m, and the estimated Compton wavelength for $H_2$ molecule is ${\displaystyle \lambda_{H}=\frac{h}{m_Hc}\simeq 6.59 \times 10^{-16}~ {\rm m}}$. The energy of the CMB photon is much smaller than the mass energy of a molecule; and the wavelength of a CMB photon is much larger that the Compton wavelength of $H_2$ molecule, so, our system satisfies the condition for ``Thompson scattering''. We can calculate the Thompson cross-section by the relation
\begin{equation}
	\sigma_T=\frac{2}{3 \pi}\left(\lambda_{H}\alpha\right)^2,
	\label{scat}
\end{equation}
where, $\alpha$ is a constant, and ${\displaystyle \alpha=\left(\frac{e}{\epsilon_0cm_H\hbar}\right)^2}$. Here, $e=1.69 \times 10^{-19}$ C, is the charge on an electron, $\epsilon_0=8.85 \times 10^{-11}~ {\rm F/m}$, is the permittivity in free space, $c=3 \times 10^{8} {\rm m/s}$, is the speed of light, and $m_H$ is the mass of single molecule of hydrogen. Replacing all the values in eq. (\ref{scat}), we get the value of scattering cross-section $\sigma_T=4.91 \times 10^{-36}~{\rm m^2}$. Now substituting $\sigma_T=\sigma_H$ in eq. (\ref{prob}), we will get the estimated value of the probability of interaction per second, $P=2.11 \times 10^{-11}~ s^{-1}$. The estimated time required for the cloud to be in equilibrium with the CMB is $t=1.5$ kyr.
	
Now we need to check the time required for the clouds to collapse. Since the Jeans time for collapse is given by the relation
\begin{equation}
t_{ff}=\sqrt{\frac{3\pi}{32G\rho_{c}}},
\end{equation}
where, $\rho_{c}=1.60 \times 10^{-18}~{\rm kgm^{-3}}$, is the central density of a pure $H_2$ cloud. Substituting the values we have the collapse time of the cloud, $t_{ff}=1.6~Myr$. Thus the cloud will achieve thermal equilibrium long before it can collapse.    
\section{\label{conclusions}Conclusion}
We see that the virial clouds are completely constrained by physical requirements, and so precisely determinable. They provide a satisfactory explanation of the observed asymmetry of the CMB Doppler shift in galactic halos. We have not only considered pure molecular hydrogen clouds, but also the possibility of these molecular hydrogen clouds, or pure interstellar dust clouds contaminated with more or less interstellar dust. We have seen that with the contamination of heavier molecules in the clouds the density of the cloud increases (see Fig.\ref{f3} and Fig.\ref{f4}), the mass and the size of the cloud decreases, since the more massive particles ``pull in'' the other molecules more (see Table-\ref{t1} and Table-\ref{t2}). The cloud mass obtained from the canonical ensemble distribution appears to be quite different form the one we obtained by using the Lane-Emden equation \cite{21}. In the virial model, the clouds have a definite boundary. 

According to  some of the observations isothermal gas spheres in the galactic halos are not stable and have no fixed boundaries with respect to the gravothermal catastrophe on time scales of a few crossing times. In general the outcome of the gravitational collapse of a gas cloud is the formation of a dense central core \cite{24}. Since, at the beginning, the evolution was almost isothermal because of the very low optical depth of the gas clouds, but, when opacity exceeded unity, the central temperature and pressure rapidly increased, which leaded to star formation. It is possible that some clouds might succeeded in avoiding this collapse and may form solid or liquid $H_2$ at less then 14 K which leaded to the formation of these dense virial clouds. In fact the evidence of the presence of liquid/solid molecules in the cold dense regions of galactic halos is present (see \cite{25,26,27}), so, this physical interpretation could help in explaining the stability of these clouds. So, we need to estimate the average optical depth of the virial clouds, $\overline{\tau}$, over a detector frequency range, $\nu_1-\nu_2$, mathematically which is given as  $\overline{\tau}=\displaystyle{\frac{1}{\nu_1-\nu_2}\int_{\nu_1}^{\nu_2}\tau_\nu~d\nu},$ to see if they are optically thick or thin \cite{28}. Instead of checking the optical depth we estimated the probability of interaction of CMB photons to check the stability time required for the virial cloud. It was seen that for pure $H_2$ clouds the interaction probability with the CMB photons; (i) when they treated as solid spheres is $\sim 1~ s^{-1}$ and the stability time was $\sim 1$ s; (ii) when the photons were treated quantum mechanically the probability $P=2.11 \times 10^{-11}~s^{-1}$ and the stability time $t= 1.5$ kyr. It is seen that the virial clouds interact with the CMB photons and come in equilibrium in almost 1.5 kyr, and the estimated time required for the cloud to collapse is 1.6 Myr. However the clouds are in gravity dominated regime they will not collapse completely, and become stable before reaching the collapse state. One can check the interaction probability for dust and the mixture of $H_2$ and dust to see how the probability and stability time could vary with varying the components of the clouds. It might be possible that the stability time could increase, since, these clouds are there for billions of years, so, this could not effect much, but it needs to be checked.

Since, the masses, sizes and densities of the clouds are different then there must be some observational aspects. One could try to estimate it and see how it could vary. As, virial clouds are at the CMB temperature, so all we would see is the CMB Doppler shift. For the M31 disk and halo, it was found that up to about $20^0$ (260 kpc) around the M31 center, in the two opposite regions of the M31 disk there was a temperature difference, $\Delta T$, of about $130 \mu$K, and it was also seen that similar effect was seen towards, the M31 halo up to 120 kpc from the M31 center with a peak value of about $40\mu$K \cite{15}. We have seen that a single virial cloud has radius is much less than $1$ pc. So, at the present level of accuracy of the Planck satellite, it is impossible to see a single cloud in the patch of the sky associated with it. All we should see is a patch of the sky with a certain number of clouds. If these clouds rotate with the the halo just like a rigid body. Then the patch of the sky with these clouds will be Doppler shifted with the halo. So, if the halo is moving towards us the patch would be red-shifted and if the halo is moving away from us the patch will be blue-shifted. This will be studied later.

It is hoped that with good models we will be able to study the dynamics of galactic halos better. We may also be able to locate some of the missing baryonic matter more precisely. However, we have not yet modeled the effect of contamination of the {\it radiation} field by higher energy radiation that would be present in the environs of the clouds.
\section*{Acknowledgements}
	We would like to acknowledge Francesco De Paolis and Achille A. Nucita, for their support and constant criticism on the work. The detailed critique of an anonymous referee, which led to an important discussion being undertaken and then inserted, is also gratefully acknowledged. AQ is grateful to IUPAP, ESA, ICRANet and AS-ICTP, for support to participate in MG15. NT is grateful to the project Inter-Asia 2018, by the collaboration of National University of Sciences and Technology, Islamabad, Pakistan and University of Salento, Lecce, Italy.

\end{document}